\documentclass[review]{elsarticle}

\usepackage[top=1.5in, bottom=1.5in, left=1.5in, right=1.5in]{geometry}
\usepackage{lineno,hyperref}
\modulolinenumbers[5]
\usepackage{graphics}
\usepackage{graphicx,epstopdf,float,wrapfig}
\usepackage{caption}
\usepackage[numbers]{natbib}
\usepackage{gensymb}
\usepackage{color}

\usepackage{epsfig}
\usepackage{amsfonts}
\usepackage{amsmath,mathtools}

\newcommand{\cutoff}{s}
\newcommand{\AIC}{{\rm AIC}}

\newcommand{\BIC}{{\rm BIC}}

\newcommand{\TR}{{\rm TR}}

\newcommand{\vh }{{h}}
\makeatletter
\def\ps@pprintTitle{%
	\let\@oddhead\@empty
	\let\@evenhead\@empty
	\let\@oddfoot\@empty
	\let\@evenfoot\@oddfoot
}
\makeatother

\biboptions{authoryear}

\begin{document}

\begin{frontmatter}

\title{Initial validation for the estimation of resting-state fMRI effective connectivity by a generalization of the correlation approach\tnoteref{t1}\vspace*{-.1in}}

\author[mainaddress]{Nan Xu\corref{correspondingauthor}}
\cortext[correspondingauthor]{Corresponding author}
\ead{nx25@cornell.edu}

\author[secondaryaddress]{R. Nathan Spreng}
\ead{nathan.spreng@gmail.com}

\author[mainaddress,thirdaddress]{Peter C. Doerschuk}
\ead{pd83@cornell.edu}

\address[mainaddress]{School of Electrical and Computer Engineering, Cornell University, Ithaca, NY, USA}
\address[secondaryaddress]{Human Neuroscience Institute, Cornell University, Ithaca, NY, USA}
\address[thirdaddress]{Nancy E. and Peter C. Meinig School of Biomedical Engineering, Cornell University, Ithaca, NY, USA}
\tnotetext[t1]{The authors certified that they do have conflict of interest.}

\begin{abstract}
Resting-state functional MRI (rs-fMRI) is widely used to noninvasively study human brain networks. Network functional connectivity is often estimated by calculating the timeseries correlation between blood-oxygen-level dependent (BOLD) signal from different regions of interest. However, standard correlation cannot characterize the direction of information flow between regions. In this paper, we introduce and test a new concept, prediction correlation, to estimate effective connectivity in functional brain networks from rs-fMRI. In this approach, the correlation between two BOLD signals is replaced by a correlation between one BOLD signal and a prediction of this signal via a causal system driven by another BOLD signal. Three validations are described: (1) Prediction correlation performed well on simulated data where the ground truth was known, and outperformed four other methods.  (2) On simulated data designed to display the ``common driver'' problem, prediction correlation did not introduce false connections between non-interacting driven ROIs. (3) On experimental data, prediction correlation recovered the previously identified network organization of human brain. Prediction correlation scales well to work with hundreds of ROIs, enabling it to assess whole brain interregional connectivity at the single subject level. These results provide an initial validation that prediction correlation can capture the direction of information flow and estimate the duration of extended temporal delays in information flow between regions of interest based on BOLD signal. This approach not only maintains the high sensitivity to network connectivity provided by the correlation analysis, but also performs well in the estimation of causal information flow in the brain. 
\end{abstract}

\begin{keyword}
	effective connectivity \sep functional connectivity \sep functional connected network \sep resting-state fMRI \sep correlation analysis 
\end{keyword}

\end{frontmatter}


\section{Introduction}\label{sec:introduction}
\par
Resting-state functional MRI (rs-fMRI) has been widely used to study the intrinsic functional architecture of the human brain based on spontaneous oscillations of the blood oxygen level dependent (BOLD) signals~\citep{smith2011network,biswal1995functional,yeo2011organization,power2011functional}. One fruitful approach has been to examine the correlations between rs-fMRI timeseries at pairs of regions of interest (ROIs) and use the correlations as a measure of connectivity strength between each pair~\citep{wig2011brainnet,sporns2011human}. The correlation method, though simple, plays a fundamental role in evaluating functional connectivity in the human brain for both task-evoked networks \citep{sadaghiani2015ongoing,cole2014intrinsic} and resting-state networks \citep{hipp2015bold,sadaghiani2015ongoing,power2013evidence}. The relationships between correlation and the topological properties, including small-world organization, modular structure, and highly connected hubs, has been studied in \cite{zalesky2012use}. However, the direction of information flow between pairs of ROIs and the causality of information flow cannot be derived from standard correlation methods. Reliable insight into the direction and causality of functional connections in the brain from BOLD signals would provide substantial breakthroughs in characterizing large-scale brain network dynamics.
\par
The BOLD signal is an indirect and sluggish measure of neuronal activity. Despite this, substantial insights have been gleaned by examining patterns of BOLD signals as proxies for functional connectivity in the brain, and these  are consistent with more direct and invasive observations \citep{Foster2015578}. At every level of analysis, the brain demonstrates an organized network structure \citep{bassett2011understanding}. So, even though neuronal activation occurs on the millisecond time scale, organized and structured activation patterns are also observed on the level of seconds, which is within the range of BOLD signals and is important for understanding cognition. Causal information about the flow of information in the brain may be detected and estimated from the BOLD signals. It remains critical, however, to evaluate methods of investigation against ground truth simulation in order to validate these methods. 
\par
Numerous methods for estimating functional or effective connectivity~\citep{van2010exploring,friston2011functional} have recently been evaluated against ground truth networks using simulated rs-fMRI data \citep{smith2011network}. Functional connectivity can be quantified with a measure of statistical dependence such as correlation, whereas effective connectivity measures the directed causal influence~\citep{friston2011functional}. In \cite{smith2011network}, performance of both types of methods across a range of measures was mixed. Standard and partial correlation excelled at detecting the presence of a connection. Other methods for estimating the direction of a connection varied from chance (Granger) to greater than 50\% accuracy (Patel's Tau and pairwise LiNGAM). These results suggest that novel methods are needed to estimate directed connectivity from rs-fMRI data, particularly with a large number of ROIs, which are necessary for full coverage of cortical and subcortical areas in the human brain. In this paper, we introduce a new method, prediction correlation, to the neuroimaging community and provide an initial validation of the approach.
\par 
Methods for estimating functional connectivity can be oriented toward estimating a real number describing strength of connectivity, which might be quite small, versus estimating a binary connectivity, which is present or absent, with possibly the addition of a strength of connectivity, in the form of a real number, for the case where a connection is present. Correlation and prediction correlation, which is a generalization of correlation that we propose in this paper, are methods that estimate a real number that describes strength of connection. Subsequent processing can then be applied to remove weak connections and/or organize the complete network into modular networks.
\par
As is described in the following sections, testing on simulated rs-fMRI data with known ground-truth networks~\citep{smith2011network} demonstrates that prediction correlation is not only sensitive in detecting network connections, as identified by standard correlation, but also achieves the highest accuracy on estimation of connection directionality among all approaches used in \cite{smith2011network} (Section \ref{sec:simulated}). In a ``common driver'' phenomena, when ROI 1 drives ROIs 2 and 3 but ROIs 2 and 3 do not directly interact, prediction correlation correctly detects strong 1$\rightarrow$2 and 1$\rightarrow$3 connections but not 2$\rightarrow$3 or 3$\rightarrow$2 connections  (Section \ref{sec:commondriver}). Finally, extending \cite{nan2014pcorrelation}, we demonstrate the robustness of this method on experimental data and that prediction correlation recovers previously identified brain network organization from experimental data (Section \ref{sec:ExperimentalPerformance}).

\section{Methods: Prediction correlation}\label{sec:p-corr}
\subsection{Fundamental method}\label{sec:p-corr_idea}
\par
In what follows, we describe a methodology for analyzing rs-fMRI data using a generalization of the well-established correlation approach, which is to correlate the timeseries at two ROIs. The generalization, denoted by ``p-correlation'' (``p'' for ``prediction'') is to replace correlation between the BOLD timeseries at two ROIs by correlation between the BOLD timeseries at one ROI and a prediction of this timeseries. The prediction is the output of a mathematical dynamical system that is driven by the timeseries at the other ROI. More generally, the prediction could be based on several, spatially discrete, ROIs. In this paper, we focus on the case where only one other ROI is used. We assume that the dynamical system is linear and has finite memory and that the memory duration and parameters may be estimated from the BOLD timeseries. If the prediction of the timeseries is restricted to use only the current  value of the timeseries that drives the dynamical system, then p-correlation is the same as standard correlation. Therefore p-correlation is a generalization of correlation. Features of p-correlation include (1) the ability  to indicate the directionality  of the interaction between two ROIs, (due to the fact that this prediction correlation is asymmetrical between two signals), and (2) the ability to evaluate the interaction based on casual information. 
\par
In the remainder of this section, we describe the p-correlation approach in detail. Consider the ordered pair of ROIs $(i,j)$ and let $x_i$ ($x_j$) denote the rs-fMRI timeseries at the $i^{\rm th}$ ($j^{\rm th}$) ROI. Both timeseries have duration $N_x$. The $x_j$ signal is predicted from the $x_i$ signal by a linear time-invariant causal dynamical model with $x_i$ as the input and the prediction $\hat x_{j|i}$ as the output. This model can be described by an impulse response, denoted $\vh_{j|i}$, which is zero for negative times. We assume that the impulse response is of finite duration, with duration denoted by $N_{h_{j|i}}$.
In summary, 
\begin{equation}
\hat{x}_{j|i}[n]=\sum_{m=0}^{N_{\vh_{j|i}}}\vh_{j|i}[m]x_i[n-m].\label{eq:LSconst2}
\end{equation}
The basic approach to estimate the coefficients of $\vh_{j|i}$ is to minimize the least squares cost \begin{equation}
\mathcal{J}(\vh_{j|i})=\sum_{n=0}^{N_x-1} (x_j[n] - \hat x_{j|i}[n])^2. \label{eq:LScost1}
\end{equation}
We estimate the value of $N_{h_{j|i}}$ and the values of the impulse response at the same time by restating the least squares problem as a Gaussian maximum likelihood estimator (MLE) with a known variance for the measurement errors. The MLE allows a trade off of the accuracy of predicting the current data (i.e., minimizing $\mathcal{J}$), which is best done by large values of $N_{h_{j|i}}$, with the accuracy of predicting when presented with new data, which is best done by smaller values of $N_{h_{j|i}}$. There are several approaches to quantifying this trade off including Akaike information criteria ($\AIC$)~\citep{Akaike1974,akaike1970statistical,sugiura1978further,hurvich1989regression,hurvich1993corrected,cavanaugh1997unifying}, Bayesian information criteria ($\BIC$)~\citep{schwarz1978estimating}, restricted maximum likelihood (REML)~\citep{thompson1962problem,patterson1971recovery}, minimum description length~\citep{rissanen1978modeling} and minimum message length~\citep{wallace1968information}. We have focused on $\AIC$ 
because it leads to easily computed problem formulations (Eq. \ref{eq:our_AIC}).
$\AIC$ realizes this balancing goal by minimizing the sum of two terms, one term that characterizes the prediction error of the dynamic system through the least squares cost $\mathcal{J}(\vh_{j|i})$ and a second term that depends on the durations $N_{h_{j|i}}$ and $N_x$: 
\begin{align}
\AIC&=
\begin{cases}
N_x\log(\frac{2\pi}{N_x-N_{h_{j|i}}}\mathcal{J}(\vh_{j|i}))+N_x+N_{h_{j|i}} & \text{if } N_x/N_{h_{j|i}}\geq 40 \\
N_x\log(\frac{2\pi}{N_x-N_{h_{j|i}}}\mathcal{J}(\vh_{j|i}))+N_x^2+\frac{N_{h_{j|i}}^2-N_x+N_{h_{j|i}}}{N_x-N_{h_{j|i}}-1} & \text{otherwise }
\end{cases}
\label{eq:our_AIC}.
\end{align}	
Simultaneous minimization of Eq.~\ref{eq:our_AIC} with respect to both $\vh_{j|i}$, which occurs only in the $\mathcal{J}(\vh_{j|i})$ term, and $N_{h_{j|i}}$ determines the duration and the value of the impulse response. The integer minimization over $N_{h_{j|i}}$ is computed by testing each value in a predetermined range of values, i.e., {1,2, ..., $D$} seconds. Then, for each value of $N_{h_{j|i}}$, the minimization with respect to $\vh_{j|i}$ involves only minimizing $\mathcal{J}(\vh_{j|i})$. Since the dynamical system describing how $x_i$ influences $x_j$ is separate from the dynamical system describing how $x_j$ influences $x_i$, the approach described here can lead to a directed rather than undirected graph of interactions between ROIs.
\par
Once $\vh_{j|i}$ and $N_{j|i}$ are estimated, the output of the dynamical system, which is the prediction $\hat{x}_{j|i}$, can be computed, and then the correlation of $x_j$ and $\hat{x}_{j|i}$, which is the so-called p-correlation, denoted by $\rho_{j|i}$, can be computed. We use ``correlation'' and $\rho_{j,i}$ for the standard approach (i.e., the standard correlation between $x_j$ and $x_i$). 
\par
\begin{figure}[H]
	\vspace{-0.3in}
		\hspace{-.5in}	\includegraphics[height=190mm]{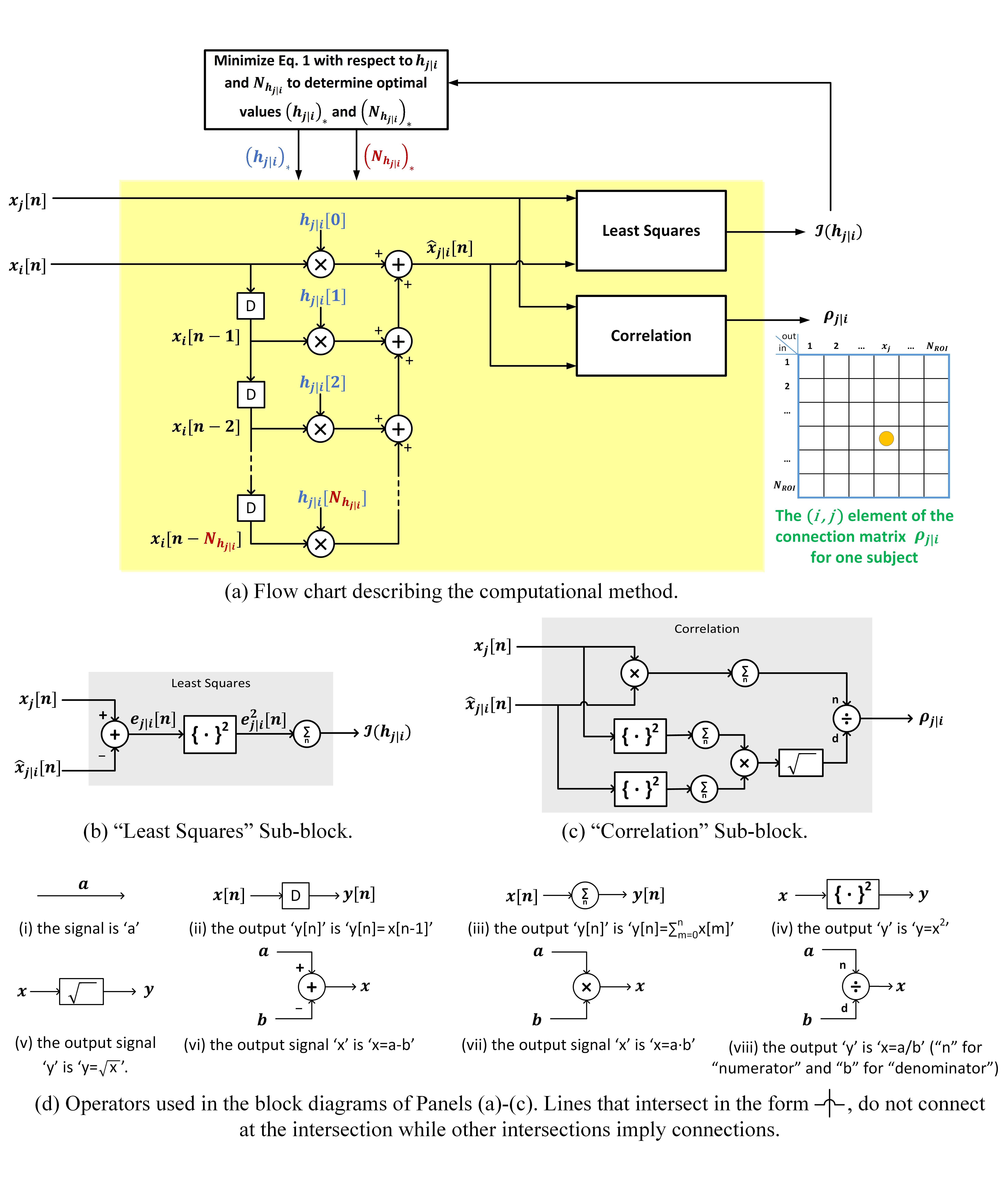}
	\caption{
		\label{fig:FlowChart}
		Block diagram and sub-block diagrams describing the computation of p-correlation for one pair of ROIs.
	}
\end{figure}

 
Let the total number of ROIs be denoted by $N_{\rm ROI}$. P-correlation is an asymmetric $N_{\rm ROI}\times N_{\rm ROI}$ matrix, where the asymmetry follows from $\rho_{j|i}\neq\rho_{i|j}$. Furthermore, p-correlation includes lags of the $x_i$ signal since the dynamical system output at time $n$, $\hat x_{j|i}[n]$, depends on the input at its current and previous times, i.e., $x_i[n]$, $x_i[n-1]$, \dots, $x_i[n-N_{h_{j|i}}+1]$. If $N_{h_{j|i}}=1$ (i.e., no lags) and $h_{j|i}[0]\geq 0$ then $\rho_{j|i}$ is the correlation between $x_j$ and $x_i$ so that $\rho_{j|i}=\rho_{j,i}$ and the approach of this paper exactly reduces to the standard approach. 
In Section \ref{sec:pcorr_constraint}, we describe a constraint such that  $h_{j|i}[0]\geq 0$ is always achieved. The p-correlation method does not depend upon the sampling rate (TR) which allows for collapsing across different scan sites or studies. The entire algorithm is shown in Fig.~\ref{fig:FlowChart}. Matlab software implementing p-correlation is available upon request.
\subsection{Specializations of the fundamental method}
In Section~\ref{sec:p-corr_idea} we defined p-correlation and described a practical method for its computation. The result is an asymmetric matrix of connection strengths for each subject. This fundamental method can be specialized for particular applications, often based on user's interests and what the user knows about the details of the applications. Several such specializations are described in the following paragraphs.  
\vspace*{-.1in}
\subsubsection{Constraints on the least squares problems}\label{sec:pcorr_constraint}
If the user has information on the type of interactions that are present, then this information can be used as a constraint on the least squares problem that determines the impulse response which is the basis of the prediction. For example, as in the simulated data of Smith et al. \citep{smith2011network}, the interactions are all positive. Constraining the impulse response values $\vh_{j|i}[n]$ to be nonnegative has implications for the values of $\rho_{j|i}$. Let $R_{j|i}$ be the covariance of $x_j$ and $\hat x_{j|i}$. $R_{j|i}$ is related to the covariance of $x_j[n]$ and $x_i[n-m]$ (i.e., the $m$-lagged covariance of the two signals, denoted by $R_{j,i}[m]$) by $R_{j|i}=\sum_{m=0}^{N_{h_{j|i}}-1}R_{j,i}[m]h_{j|i}[m]$. The covariance $R_{j|i}$ is the numerator of $\rho_{j|i}$. Therefore, if all the lagged covariances are positive and we require the estimated values of $h_{j|i}[m]$ to be positive then we are assured of getting a nonnegative value for $R_{j|i}$ and for the p-correlation $\rho_{j|i}$. In the traditional functional connectivity analysis, when global signal regression is applied to rs-fMRI timeseries data, the valid inference of negative correlations cannot be made \citep{journals/neuroimage/MurphyBHJB09,journals/brain/SaadGMCJMC12}, and only positive correlations are interpreted. In this situation, the nonnegative ``constrained'' estimation approach is appropriate.
\vspace*{-.1in}
\subsubsection{Thresholding $\rho_{j|i}$}\label{sec:pCorr_threshold}
\par 
Three natural methods for thresholding $\rho_{j|i}$ are described in this section.  
\par 
Even with $h_{j|i}[n]\ge 0$, it may be that p-correlation is not positive because one or more of the $m$-lagged covariance values are negative. Therefore, if non-negativity is required, we replace all negative $\rho_{j|i}$ values by zeros. 
One reason for seeking to have $\rho_{j|i}$ non negative is mean signal regression in the preprocessing of the fMRI data which makes it difficult to interpret negative correlations. However, alternative preprocessing which omits mean signal regression~\citep{journals/jam/JoGRBMCS13} removes this requirement.
\par 
The previous paragraph concerned thresholding at value 0. Higher data-dependent minimum thresholds are often used for correlation and the same approach can be applied to p-correlaton. A standard approach \citep{power2011functional} is to order the values of correlation and leave the top $s$ percent of values unchanged and set the remaining values to zero. In other words, the threshold $\gamma(\cutoff)$ is set to be the 100-$\cutoff$ percentile of all values in the p-correlation matrix. 
\par 
In some problems the interactions are known to be unidirectional, e.g., in the simulated data of Smith \citep{smith2011network}. In this situation, a third thresholding method, which makes p-correlation unidirectional, is natural. The threshold is to consider the two transpose-related elements of the matrix and set the smaller to zero and leave the larger unchanged.
\par
Each of the thresholding methods is a nonlinear operation applied to the matrix of $\rho_{j|i}$ coefficients. Each can be applied to any matrix $M$ to give an output matrix $N$, in particular, in the order of the previous three paragraphs,
\begin{subequations}
	\label{eq:Threshold}
	\begin{align}
	& N_{ij}=\begin{cases}M_{ij}, & \mbox{if } M_{ij}\geq 0\\0, & \mbox{otherwise}\end{cases}, \label{eq:Threshold_pos}\\
	& N_{ij}=\begin{cases}M_{ij}, & \mbox{if }M_{ij}\geq \gamma(s)\\0, & \mbox{otherwise}\end{cases},\label{eq:Threshold_tops}\\\nonumber
	&\mbox{where $\gamma(s)$ is the $100-\cutoff$ percentile of all values in $M$, and }\\
	& N_{ij}=\begin{cases}M_{ij}, & \mbox{if } M_{ij}\geq M_{ji}\\0, & \mbox{otherwise}\end{cases}. \label{eq:Threshold_dir}
	\end{align}
\end{subequations}
The thresholding approach forms a $N_{\rm ROI}\times N_{\rm ROI}$ matrix of thresholded connection weights, from which the network is computed. 
\vspace*{-.1in}
\subsubsection{Averaging over subjects}\label{sec:pCorr-average}
Some investigations, e.g.,~\cite{smith2011network,laumann2015functional}, are interested in estimating subject-by-subject details, but in many other investigations on functional networks of human brain using experimental data, e.g.,~\cite{power2011functional,power2013evidence,schaefer2014dynamic,gordon2016generation}, there is averaging over subjects in order to improve the SNR. Just as the thresholding methods (Section \ref{sec:pCorr_threshold}), which are nonlinearities that can be applied to any matrix, the averaging we use can be applied to any family of matrices $M_k$ ($k\in\{1,\dots, K\}$, where $K$ is the number of subjects) to give an output matrix $N$ via $N=\frac{1}{K}\sum_{k=1}^{K}M_k$. The functional network estimated by the averaged p-correlation matrix can be further clustered into sub-networks through a graphic theoretic analysis. 
\vspace*{-.1in}
\subsection{Extension to multi-subject processing}
There is a recent interest in estimating effective networks from multiple subjects while accommodating the heterogeneity of the group~\citep{smith2012future,ramsey2010six,gates2012group}. Specifically, the IMaGES algorithm~\citep{ramsey2010six} estimates one generalized network from a group by assuming all subjects are homogeneous, and the GIMME algorithm~\citep{gates2012group} can further refine the estimate for each individual subject from the general information estimated from the whole group. IMaGES and GIMME are based on existing single-subject methods, specifically GES for IMaGES and uSEM and euSEM for GIMME and, when applied to groups of appropriate size, both GIMME and IMaGES provide more accurate estimates of effective connectivity than the single subject methods on which they are based~\citep{ramsey2011multi,gates2012group}.
\par
Information concerning groups of subjects could also be used in p-correlation. One approach would be to replace the $\vh_{j|i}$ in Eq.~\ref{eq:LSconst2} by $\vh_{j|i}^g+\vh_{j|i}^l$, where $\vh_{j|i}^g$ is the group component common to all subjects, and $\vh_{j|i}^l$ is the component unique to the specific subject $l$. In this approach, Eq.~\ref{eq:LSconst2} would be generalized to
\begin{equation}
\hat{x}_{j|i}^l[n]=\sum_{m=0}^{N_{\vh_{j|i}^g}}\vh_{j|i}^g[m]x_i^{g}[n-m]+\sum_{k=0}^{N_{\vh_{j|i}^l}}\vh_{j|i}^l[k]x_i^l[n-k]^l\label{eq:LSconst2_g}
\end{equation}
where $N_{\vh_{j|i}^g}$ and $N_{\vh_{j|i}^l}$ are the probably different durations of the two components of the causal finite-duration impulse response. There are two issues when using Eq.~\ref{eq:LSconst2_g}. First the $\AIC$ analysis must be generalized in order to determine two impulse response durations where one is common to the entire group of subjects. Second, in order to require the least squares to use the group impulse response and not just set it to zero, a regularizer such as $\sum_{m=0}^{N_{\vh_{j|i}^l}}(\vh_{j|i}^l[m])^2$ must be added to the least squares cost. While both of these issues can be addressed, in the current paper, we only focus on the individual analysis, which may be the only meaningful option under certain circumstances, i.e., a clinical environment.

\section{Results}
\subsection{Application on Simulated Data}\label{sec:simulated}
\subsubsection{Data source: simulated BOLD timeseries}
\label{sec:Smith-participants}
Simulated fMRI timeseries from the laboratory of S. M. Smith are documented~\citep{smith2011network} and available on-line (\url{http://www.fmrib.ox.ac.uk/analysis/netsim/}). These timeseries have been used as benchmark simulated fMRI data for testing effective connectivity~\citep{smith2011network,gates2012group,ramsey2011multi,hyvarinen2013pairwise}. The simulations are based on a variety of underlying networks of different complexity and can be described as having three levels. First there is a neural level which is a stochastic linear vector differential equation which produces a neural timeseries for each ROI. Second, for each ROI, there is a nonlinear balloon model driven by the corresponding neural timeseries which produces a vascular timeseries. Third, for each ROI, the fMRI timeseries is the vascular timeseries plus thermal noise. To simulate preprocessing of fMRI data, a highpass filtered at a cutoff frequency of 1/200s was applied to each simulation (most recently revised on Aug. 24, 2012 based on the website \url{www.fmrib.ox.ac.uk/analysis/netsim}). The current paper considers the first four sets of simulations from \cite{smith2011network}, $Sim1-Sim4$, which 
are the four most ``typical'' network scenarios provided in \cite{smith2011network}, and which are based on different underlying networks with sizes 5, 10, 15, and 50 ROIs, respectively. 
\par
These synthetic fMRI timeseries were sampled every 3s ({\TR}$=3s$) and the total duration is $N_x=10$~mins. All four simulations have $1\%$ thermal noise and the hemodynamic response function (HRF) used in the second step has standard deviation of $0.5$~s. The simulation is repeated for each of 50 subjects. 
\subsubsection{Specialization on p-correlation for the processing of the simulated data}
\label{sec:simulated:special}
\begin{figure}[H]
	\centering\includegraphics[height=50mm]{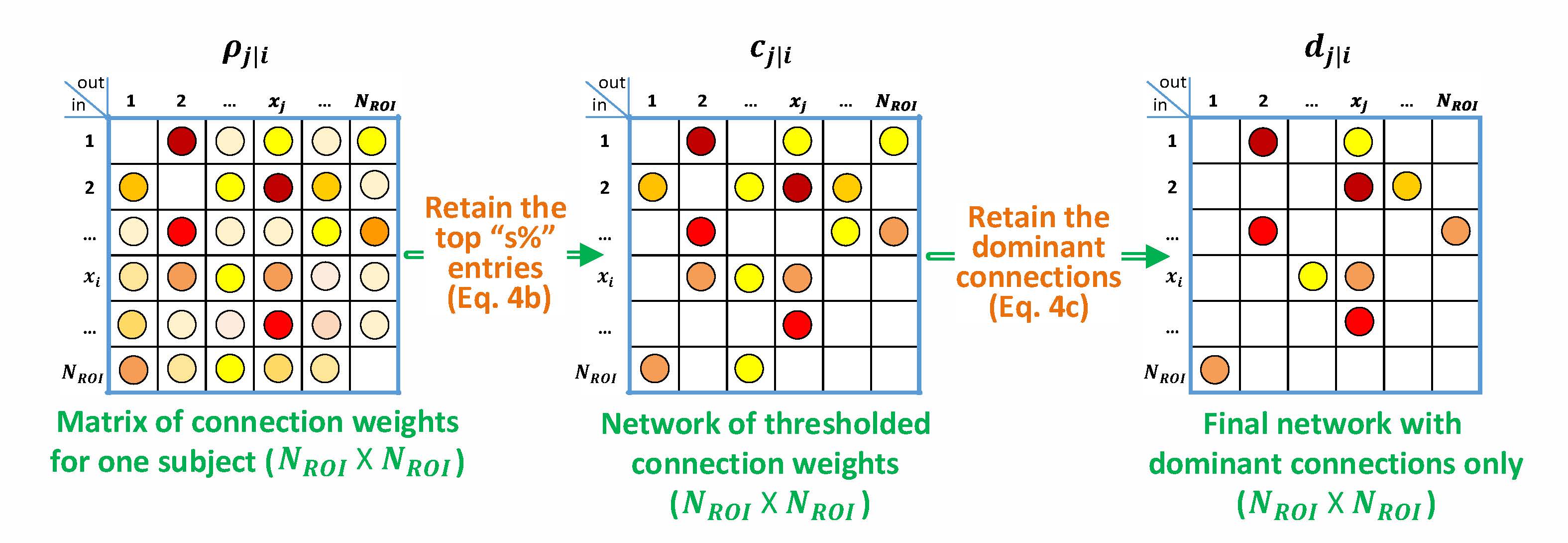}
	\vspace*{-.0in}
	\caption{
		\label{fig:special_sim}
		Block diagram describing the specialization of p-correlation for simulated data. Nonzero entries are filled by colored dots with higher values represented by ``hotter'' colors and lower values represented by ``colder'' colors, and zero entries are left as blank in the above matrices.
	}\vspace*{-.1in}
\end{figure}
\par 
The algorithm is shown in Figure \ref{fig:special_sim}. Given that the interactions are all positive in the simulated data, it is natural to apply the nonnegative constraint on the least squares problem so that no negative impulse responses are allowed. Although unconstrained p-correlation is also computed on the simulated data, looking forward to Section \ref{sec:restuls}, the numerical results indicate that the constrained version is more appropriate.
\par 
As is described above, the integer minimization over the impulse function duration, $N_{h_{j|i}}$, is computed by testing from 1 second up to $D$ seconds. Assuming that knowledge of the behavior of a ROI over the past 15 seconds is sufficient to describe its effect on a second ROI, we restricted the temporal window for directional influence between ROIs to no more than 15s, i.e. $D=15s$.
\par
\begin{figure}[H]
	\begin{center}		
		\includegraphics[height=0.21\textwidth]{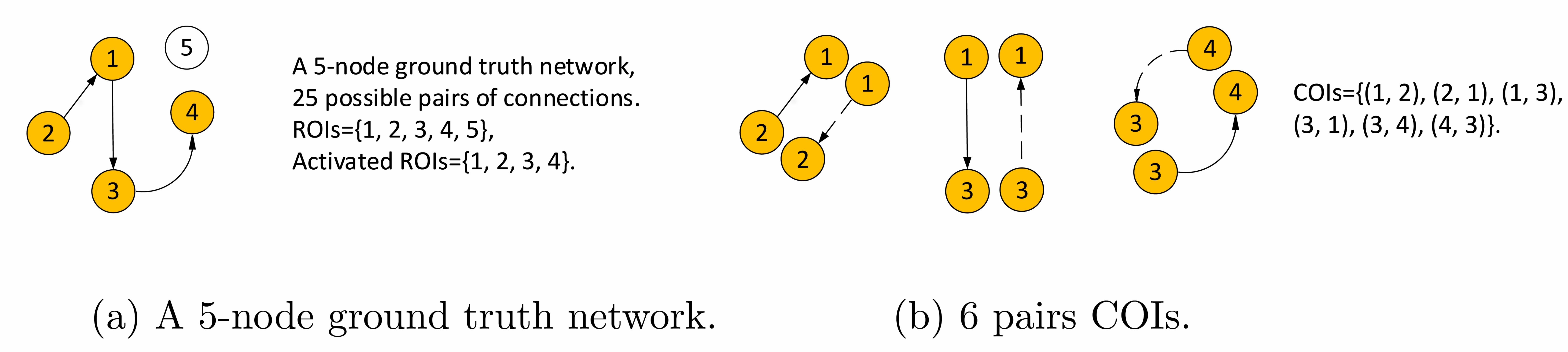}
	\end{center}
	\caption{
		\label{fig:5nNetworkw}
		Example calculation of the threshold $\cutoff$ for a 5-node network. (a) The network with activated ROIs shown in orange. The number of all possible connections is $5^2=25$. (b) The 6 COIs, where the dashed lines are connections that do not existed in the ground truth but still are considered interesting. Therefore, $\cutoff=6/25$ for this network.
	}
\end{figure}

Next, we consider the choice of threshold, $\cutoff$ in Eq. \ref{eq:Threshold_tops}. We use this method in order to exploit all of the a priori knowledge about the simulated data. Since the underlying ground truth networks for the simulated fMRI timeseries, denoted by $a_{j|i}$, are given, the threshold value $\cutoff$ is among our prior knowledge as is described below. We denote ROIs that are involved in the connections of the ground truth network as active ROIs. All connections involving the active ROIs are connections of interest (COIs), including connections that are actually absent such as the reverse connection in an unidirectional interaction. The value of $\cutoff$ is then the ratio of the number of COIs and the number of all possible connections, which gives $\cutoff=40,~22,~16$ and $4$ percent for the four simulations, respectively. An example of computing $\cutoff$ for a 5-node network is shown in Fig. \ref{fig:5nNetworkw}.
For the Smith simulated data, we have additional prior knowledge that the networks contain only unidirectional connections. Therefore, as is also done in~\cite{smith2011network}, we compare our estimated network $d_{j|i}$, which includes the unidirectional condition, with the ground truth network $a_{j|i}$. The estimated network $d_{j|i}$ is the output of Eq. \ref{eq:Threshold_dir} where the input is the thresholded network $c_{j|i}$.
\subsubsection{Performance Criteria}\label{sec:simulated:accuracy}
\par
To compare the computed and ground truth networks, we define ``accuracy", denoted by $\mathcal{A}$. In particular, $\mathcal{A}$ is defined to be the mean fractional rate of detecting the correct directionality of true connections. Specifically, it is defined to be
\begin{equation}
\mathcal{A}=\frac{\sum_{i=1}^{N_{\rm ROI}}\sum_{j=1}^{N_{\rm ROI}} 1\{a_{j|i}>0\}1\{d_{j|i}>0\}}{\sum_{i=1}^{N_{\rm ROI}}\sum_{j=1}^{N_{\rm ROI}} 1\{a_{j|i}>0\}},\label{eqn:accuracy}
\end{equation}
where $1\{L\}$ is 1 if $L$ is true, and 0 otherwise. Like the computation of the ``d-accuracy'' introduced in~\cite{smith2011network}, $\mathcal{A}$ evaluates the percentage of the correct directionality ($\mathcal{A}$ is between 0 and 1). The threshold operation introduced above (Section \ref{sec:simulated:special}) differentiates the performance of directional analytical methods based on their sensitivity. The more sensitive the method is, the more true connections it can detect.
Notice that application of the threshold $\cutoff$ leads to $d_{j|i}$ values that are almost certainly far from zero or exactly zero. Computing the accuracy $\mathcal{A}$ after the threshold operation tells the directionality after knowing the presence of the connections, which enables us to evaluate the overall performance of sensitivity and directionality of a directional analytical method.

\subsubsection{Alternative methods for effective networks estimation}\label{sec:simulated:competitors}
\par 
P-correlation and four alternative methods from~\cite{smith2011network}, specifically,  ``Granger B1'', ``Gen Synch S1'', ``LiNGAM'' and ``Patel's conditional dependence measure'', were compared by the accuracy criteria ($\mathcal{A}$), since under both synthetic and experimental scenarios, these methods have been tested and have relatively good performances among all the others~\citep{smith2011network,dawson2013evaluation}. The computation of these methods were done by software provided by Prof. S.M. Smith. Granger B1, a pairwise Granger causality estimation method which provides the best performance among Granger causality approaches~\citep{smith2011network,dawson2013evaluation}, uses the Bayesian Information Criterion to estimate the lag up to 1 {\TR}. Gen Synch S1 is a nonlinear synchronization method with respect to the time lag 1 {\TR}. It ``evaluates synchrony by analyzing the interdependence between the signals in a state space reconstructed domain"~\citep[p.~671]{dauwels2010comparative}. The LiNGAM (Linear, Non-Gaussian, Acyclic causal Models) algorithm is a global network model utilizing higher-order distributional statistics, via independent component analysis, to estimate the network connections. Patel's conditional dependence measure investigates the causality from the imbalance between two conditional probabilities, $P(x_j|x_i)$ and $P(x_i|x_j)$. P-correlation, Granger B1, Gen Synch S1 and LiNGAM all compute an asymmetric matrix filled with real-number connection weights, analogous to our $c_{j|i}$. In all cases, the unidirectional prior knowledge is applied analogous to our transformation from $c_{j|i}$ to $d_{j|i}$. 
For the Patel method implemented by~\cite{smith2011network}, the thresholding operation was applied on ``Patel's $\kappa$ bin 0.75'' matrix, while the directionality was determined by ``Patel's $\tau$ bin 0.75'' matrix. 
\par
In addition to the algorithms included in~\cite{smith2011network}, IMaGES~\citep{ramsey2010six} and uSEM~\citep{kim2007unified} which is the estimation method for resting-state fMRI employed by GIMME algorithm, have also been tested on the same set of simulated data \citep{ramsey2011multi,gates2012group}. Results reported in \cite{ramsey2011multi,gates2012group} show that their estimation based on the single subject is either similar to or less good than the best-performing method provided in \cite{smith2011network}. 
\par
Comparing p-correlation with alternative methods of estimating effective connectivity, p-correlation provides a full asymmetric matrix for each subject independent of all other subjects, in which each entry, like correlation, predicts a connection strength between two ROIs. The ability to compute results based on an individual subject means that p-correlation can potentially be used in a clinical environment. This full asymmetric matrix of p-correlations can be thresholded as desired and/or further processed as desired using another algorithm, i.e., a graph analytic algorithm. In addition, p-correlation can process networks with hundreds of ROIs while GIMME is limited to 3-25 ROIs (Page 3 of GIMME Manual (Version 12)). Furthermore, p-correlation estimates the temporal causal relation in the form of lagged impulse response in addition to the spatial causal relation between any pair of ROIs. In contrast, some alternative algorithms (e.g. IMaGES) estimate a sparse graph of interactions, and thus solve a somewhat different problem than the p-correlation method. Other algorithms have been developed as post-processing algorithms, which cannot detect connections, but only estimate direction if connections are detected by other methods, e.g., correlation. Among them, pairwise LiNGAM~\citep{hyvarinen2013pairwise} achieved success on Smith's data~\citep{smith2011network}. Several algorithms, such as Patel's $\tau$, LiNGAM and pairwise LiNGAM, chose one of the two possible directions for each pair of ROIs. Such unidirectionality may be appropriate in some situations. Alternative algorithms, including p-correlation, provide strengths for both directions, where the two strengths may be quite different when one direction is dominant.

\subsubsection{Results on simulated data}\label{sec:restuls}
The methods described in this paper were implemented in Matlab software, which is available upon request, and were applied to four of Smith's fMRI simulations~\citep{smith2011network}. The four simulations are $Sim1-Sim4$ which have a variable number of ROIs (5, 10, 15, 50) but no confounding variables.
\par
The p-correlation method is based on estimation of a linear time-invariant causal dynamic model. The sample means of the duration of either constrained or unconstrained impulse responses are  3.34s, 3.58s, 3.64s and 3.76s for the 4 simulations, respectively. By limiting the impulse response duration to 1 {\TR}, it was verified that p-correlation with constraint on Least Squares is equivalent to the standard correlation as is described in Section \ref{sec:introduction}. After thresholding the p-correlations computed with the nonnegative constraint on the coefficients of the linear system, an asymmetric matrix of connection weights $c_{j|i}$ for each subject was obtained.
\par
\begin{figure}[H]
	\hspace*{.5cm}
	\centerline{	\includegraphics[height=28.5mm]{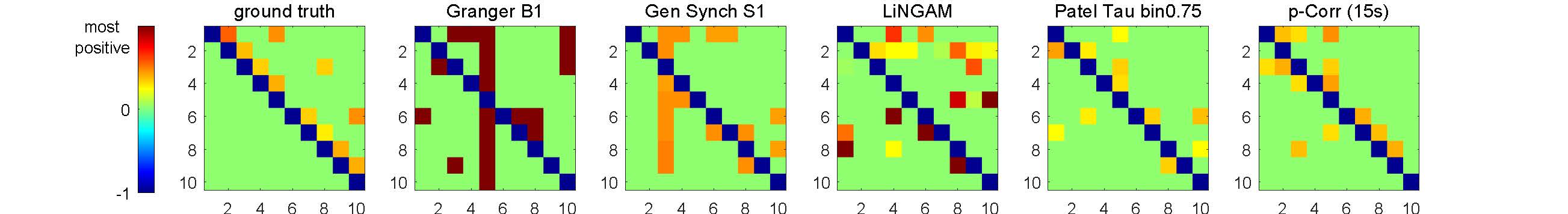} }
	\caption{
		\label{fig:NetworkComparison}
		Images of $a_{j|i}$ (for ground truth) and $d_{j|i}$ (for constrained p-correlation), and quantities analogous to $d_{j|i}$ (for Granger B1, Gen Synch S1, LiNGAM, and Patel) for Subject 14 of $Sim$2. Each image uses the same ordering of colors, but has different range of numerical values.
	}
\end{figure}
\begin{figure}[H]
	\hspace*{.05in}
	\centerline{\includegraphics[height=80mm]{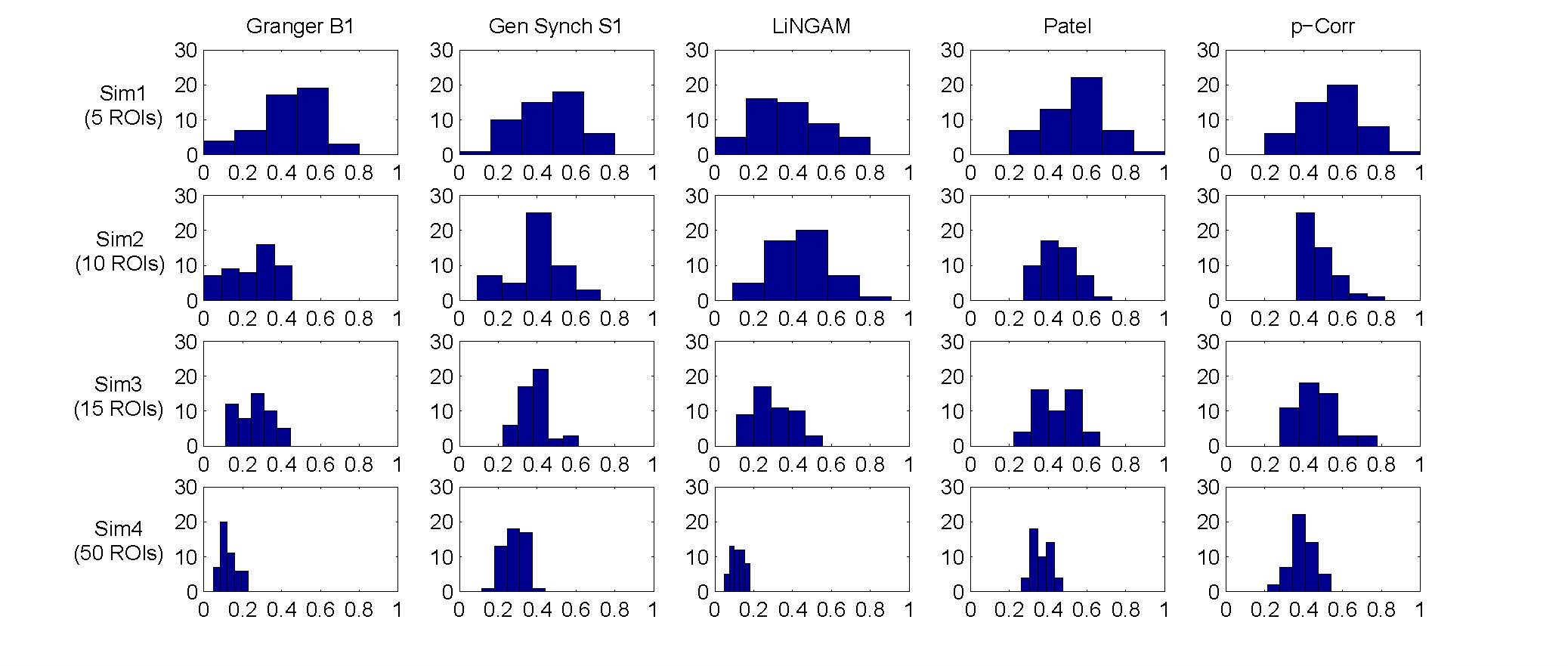} }
	\vspace*{-.3in}
	\caption{
		\label{fig:Histogram}
		Accuracy histogram for Granger B1, Gen Synch S1, LiNGAM, Patel and constrained p-correlation.
	}
\end{figure}

The same specifications for processing of the simulated data, in particular, the same choice of the $\cutoff$ threshold (Eq.~\ref{eq:Threshold_tops}) and the knowledge of unidirectionality (Eq.~\ref{eq:Threshold_dir}), have also been applied to the results of four alternative methods introduced in Section \ref{sec:simulated:competitors}. The performance of all five methods was evaluated by the accuracy criteria $\mathcal{A}$ (Eq. \ref{eqn:accuracy}) for each subject. Fig. \ref{fig:NetworkComparison} shows the input to the accuracy criteria $\mathcal{A}$, i.e., $a_{j|i}$ and $d_{j|i}$, for Subject 14 of $Sim$2.
The mean and standard deviation of accuracy for each simulation, i.e., the average and square root of the sample variance of $\mathcal{A}$ (Eq. \ref{eqn:accuracy}) over all 50 subjects, were computed and the results are tabulated in Table \ref{tab:DirectionCorrect}. For all four simulations, constrained p-correlation achieved the highest accuracy compared to other methods. The unconstrained p-correlation is less appropriate when applied to a network with all positive connection weights. We also computed the mean and standard deviation of $\mathcal{A}$ for pairwise LiNGAM, which gives .566$\pm$.138, .656$\pm$.206, .510$\pm$.119 and .506$\pm$.056 for four simulations, respectively. The result shows the highly accurate directionality that pairwise LiNGAM can achieve in this particular unidirectional network setting. Histograms displaying the distribution of accuracy for the five methods for each simulation are shown in Fig. \ref{fig:Histogram}. The superior performance of p-correlation is demonstrated by the fact that the bulk of the histogram is further to the right, and the left tail is less massive.
\begin{table}[h]
	\begin{center}
		\hspace*{-.5in}\begin{tabular}{|l|c|c|c|c|c|c|c|}
			\hline
			{\footnotesize Simulation} & 1 & 2 & 3 & 4 \\\hline
			{\footnotesize \# of ROIs} & 5 & 10 & 15 & 50   \\\hline
			{\footnotesize \# of COI pairs} & 10 & 22 & 36 & 122  \\\hline
			{\footnotesize Granger B1} &.440$\pm$.206 &.295$\pm$.127 &.262$\pm$.088 &.130$\pm$.044\\\hline
			{\footnotesize Gen Synch S1} &.472$\pm$.201 &.405$\pm$.139 &.379$\pm$.079 &.285$\pm$.056\\\hline
			{\footnotesize LiNGAM}	&.372$\pm$.229 &.435$\pm$.177 &.301$\pm$.106 &.119$\pm$.037\\\hline
			{\footnotesize Patel}   &.528$\pm$.193	 &.491$\pm$.101 &.446$\pm$.099 &.366$\pm$.048\\\hline
			{\footnotesize p-Corr (constrained)}  &.532$\pm$.192 &.502$\pm$.114 &.457$\pm$.126 &.405$\pm$.065\\\hline
			{\footnotesize p-Corr (unconstrained)} &.520$\pm$.218 &.467$\pm$.123 &.439$\pm$.109 &.371$\pm$.058\\\hline			
		\end{tabular}
	\end{center}
	\caption{
		\label{tab:DirectionCorrect} 
		Comparison of the mean and standard deviation of accuracy over 50 subjects among different methods. The standard deviation tend to be larger on the smaller netowrks ($Sim$1-$Sim$4 have 5, 10, 15, 50 ROIs, respectively) because one error is proportionally of larger impact in a smaller network.
	}
\end{table}

{
	\subsection{The performance of correlation and p-correlation on common drivers}\label{sec:commondriver}
	A ``common driver'' situation is the case where ROI 1 drives ROIs 2 and 3 but ROIs 2 and 3 do not directly interact. The challenge is to correctly detect the 1$\rightarrow$2 and 1$\rightarrow$3 connections without detecting 2$\rightarrow$3 or 3$\rightarrow$2 false connections. In order to focus exclusively on this situation, we have computed synthetic data from the three-ROI network shown in Fig. \ref{fig:commondriver} and defined by 
	\begin{align}
	x_{1}[n+1]&=a_1x_1[n]+b_1w_1[n]\label{eq:commondriver1}\\
	x_2[n+1]&=a_2x_2[n]+a_{21}x_1[n]+b_2w_2[n]\label{eq:commondriver2}\\
	x_3[n+1]&=a_3x_3[n]+a_{31}x_1[n]+b_3w_3[n]\label{eq:commondriver3}
	\end{align}
	\begin{figure}[H]
		\centering\includegraphics[height=40mm]{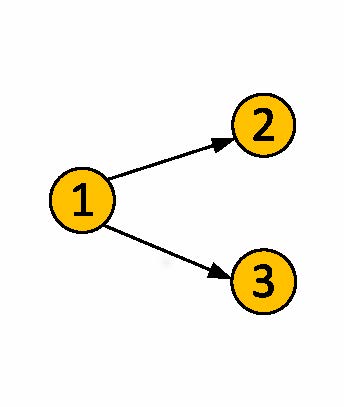}
		\caption{
			\label{fig:commondriver}
			The common driver problem.
		}
	\end{figure}
	where $w[n]=[w_1[n], w_2[n], w_3[n]]^T$ is an independent and identically distributed Gaussian stochastic process with mean 0 and variance $I_3$ (the  3$\times$3 identity matrix). \cite{zalesky2012use} consider mathematical models of this type and give theoretical results for correlations. The system is initialized in the steady state and simulated for 1000 steps, $N_x=1000$. We consider only $a_1=a_2=a_3=.8$ (so that all ROIs have the same intrinsic memory duration) and $b_1=b_2=b_3=.2$ (so that all ROIs have the same intrinsic noise power, and the intrinsic noises are all independent). We consider the following cases:
	(1) no driving: $a_{21}=a_{31}=0$, 
	(2) weak driving: $a_{21}=a_{31}=.1$, 
	(3) strong driving: $a_{21}=a_{31}=.4$, and  
	(4) asymmetrical strong driving: $a_{21}=.4$, and $a_{31}=.1$.
	
	Each simulation was repeated for 50 subjects. Let the maximum allowable duration of the impulse response be 3 samples. By using the specialization of p-correlation for Smith simulated data, as is described in Section \ref{sec:simulated:special}, a directed graph $d_{j|i}$ is estimated by p-correlation (Fig. \ref{fig:special_sim}) and the correlation matrix is computed for each subject. The steady state covariance of Eqs. \ref{eq:commondriver1}-\ref{eq:commondriver3} is the correlation matrix. 
	In Case (1), the mean and standard deviation of nonzero entries of $\rho_{j|i}$ with constrained least squares (Section \ref{sec:pcorr_constraint}) are 5.384e-04$\pm$0.072. This number becomes 0.058$\pm$0.043 when unconstrained least squares is applied. The smaller magnitude of the results using constrained least squares indicates that taking advantage of the prior knowledge that the weights are positive (i.e., $a_1=a_2=a_3=.8$) provides improved performance in this case. In Cases (2) and (3), both the constrained and the unconstrained least squares achieve a 100\% accuracy (Eq. \ref{eqn:accuracy}) for each subject. In the fourth case, the constrained or the unconstrained least squares gives an average of .800$\pm$.247 accuracy over all 50 subjects. We also tested $N_x=$200, 500, 5000 for all four cases. Notice that as $N_x$ goes large, correlations become closer to the steady state and the accuracy computed by the p-correlation method increases as well.
	\par 
	In addition, p-correlation estimated the correct hierarchy on the three pairs of connection weights, which are consistent with ``strong'', ``weak'' and ``non-'' connections in the ground truth network. It also shows the correct direction of connections in a pair by a stronger weight. The constrained least squares (Section \ref{sec:pcorr_constraint}) provides a slightly superior result than the unconstrained approach. Specifically, larger numerical differences between the zero and nonzero entries, as well as between the asymmetric strong weights, were shown. On average across all 50 subjects, p-correlation used an impulse response duration of 1.007 samples for all four cases for both constrained and unconstrained approaches. In addition, in Case (3) (asymmetric strong weights), correlation mis-detected the connection between node 2 and 3, specifically the 2-3 correlation was the highest correlation value among the three pairs, whereas p-correlation, for both the constrained and unconstrained approaches, estimated this value as the lowest of the three pairs thereby avoiding the error in the correlation results.
}
\subsection{Performance on experimental fMRI data}\label{sec:ExperimentalPerformance}
\begin{figure}[H]
	\hspace{-0.7in}\includegraphics[height=42mm]{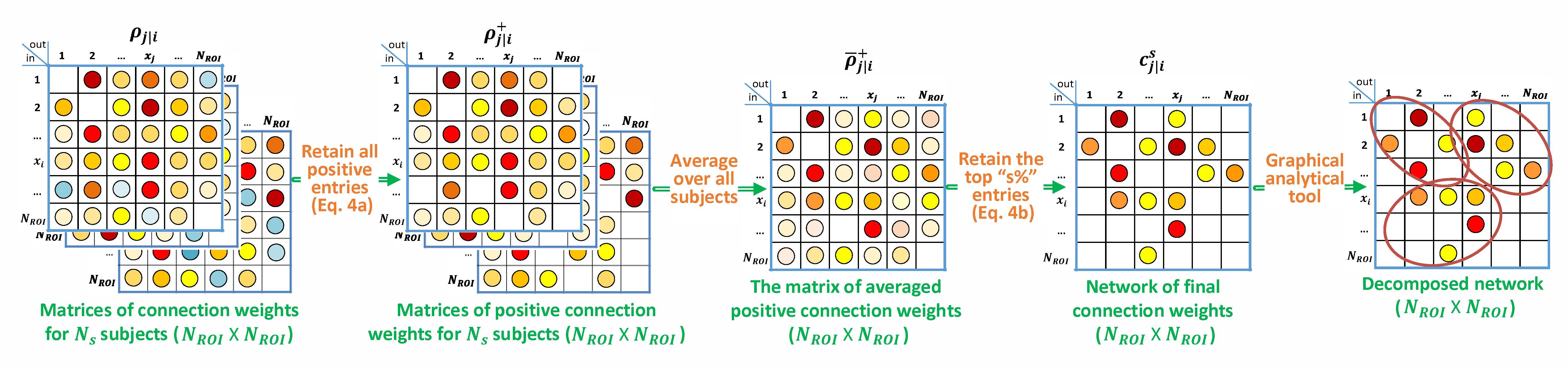}
	\caption{
		\label{fig:special_exp}
		Block diagram describing the specialization of p-correlation for the experimental data. Nonzero entries are filled by colored dots with higher values represented by ``hotter" colors and lower values represented by ``colder" colors, and zero entries are left as blank in the above matrices.
	}
\end{figure}
While the tools described in this paper can be assembled into many algorithms, we use only one algorithm, which is shown in Figure \ref{fig:special_exp}, to further characterize \citep{nan2014pcorrelation}, a cohort of 132 subjects from the 1000 Functional Connectomes Project (\url{http://www.nitrc.org/projects/fcon_1000/})~\citep{biswal2010toward}. This data is provided from different scanning sites, and thus has variable sampling rates (TR’s = approx 1-3s, mean $\pm$ standard deviation of 2.3 $\pm$ 0.4s). The scan duration also varied from 119-295 TRs (mean ± standard deviation of 167.5 $\pm$ 41.7). The data from the whole brain were preprocessed~\citep{anderson2011network}, linearly detrended and bandpass filtered (retaining signal between 0.001 and 0.1 Hz), and motion scrubbed~\citep{power2012spurious} with the threshold set to 0.2. The preprocessed rs-fMRI BOLD signal was extracted from $N_{\rm ROI}$ = 264 spherical ROIs each with a 10mm diameter. We combine our p-correlation ideas with the widely-used~\citep{power2011functional,power2012spurious,lahnakoski2012naturalistic,gordon2016generation} Infomap graph analytical algorithm~\citep{PhysRevE.80.056117} to determine networks within the set of 264 ROIs.
\par
\begin{figure}[H]
	\hspace*{.05in}
	\centerline{\includegraphics[height=50mm]{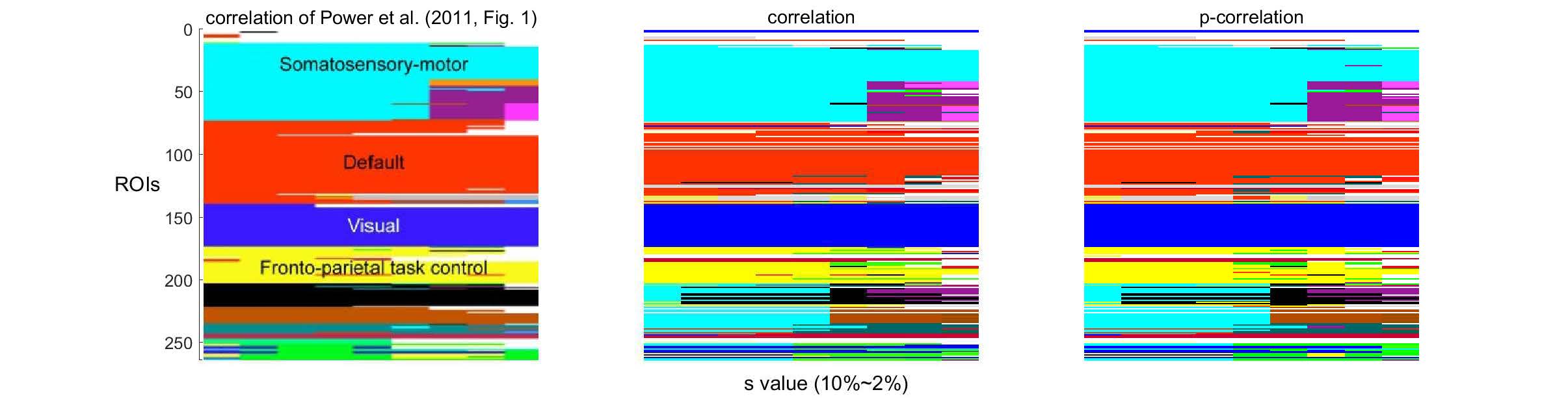}}
	\vspace*{-.3in}
	\caption{
		\label{fig:Stability_BIC}
		The stability of networks across various thresholding criteria ($s$). The white regions indicate ROIs that belong to networks with less than four ROIs.
	}
\end{figure}

 
As a function of the value of the threshold $\cutoff$, Infomap creates a variable number of networks. Following~\citet[Fig.~1]{power2011functional}, the network stability over a range of threshold $\cutoff\in\{2,\dots,10\}$ using correlations and p-correlations are shown in Fig.~\ref{fig:Stability_BIC}, in which different networks are represented by different colors. Similar to~\cite{power2011functional} (the first figure in Fig \ref{fig:Stability_BIC}), we note that the assignment of ROIs to networks remains relatively constant over all values of the threshold $\cutoff$, illustrated by the constant horizontal bands in different colors. Also, networks are hierarchically refined as $\cutoff$ rises. In summary, the number of networks increases as the value of $\cutoff$ decreases, and p-correlation replicated the brain network organizations that were detected by correlation. The network results are consistent with the network organizations detected in~\cite{power2011functional}.
\par
\begin{figure}[H]
	\begin{center}
		\includegraphics[width=12cm]{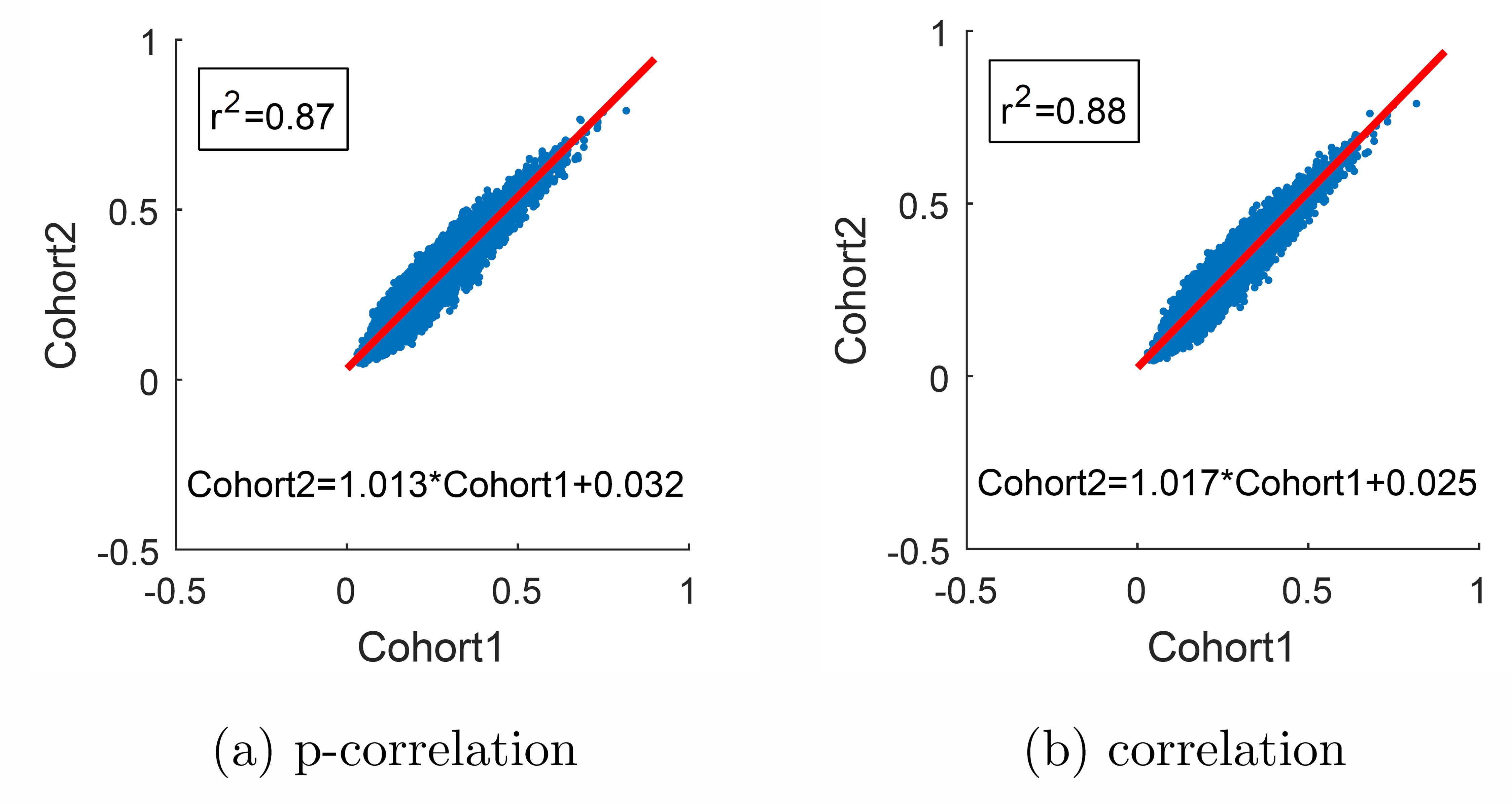}
	\end{center}
	\caption{\label{fig:Scatter_corr_pcorr}Scatter plot of p-correlation and correlation for the two cohorts. The red line is the Least Squares fit for predicting Cohort 2 from Cohort 1. Only positive values are used in the Least Squares calculation and shown in the plot.}
\end{figure}

In order to test the robustness of the p-correlation calculation, all 132 subjects were randomly divided into two equal cohorts, and each cohort was separately processed. The average of p-correlation connection strength ${\rho}_{j|i}^+$ across all subjects in the cohort, which is denoted by $\bar{\rho}_{j|i}^+$, is shown as a scatter plot in Fig. \ref{fig:Scatter_corr_pcorr} (a) (in Fig. \ref{fig:Scatter_corr_pcorr}, all (0,0) points are removed). The linear least squares prediction of Cohort 2 from Cohort 1 is a close fit to the data ($r^2=.87$) and is nearly a 45$^{\circ}$ diagonal line ($\bar{\rho}_{j|i}^\text{Cohort 2}$=1.013$\bar{\rho}_{j|i}^\text{Cohort 1}+$.032), thereby indicating the robust nature of p-correlation. Following the same procedure, the average of correlation connection strength ${\rho}_{i,j}^+$ across all subjects in the cohort, which is denoted by $\bar{\rho}_{i,j}^+$, is shown in Fig. \ref{fig:Scatter_corr_pcorr} (b). Comparing Fig.\ref{fig:Scatter_corr_pcorr} (a) and (b) indicates that the p-correlation achieves the same robustness as correlation. Additional plots in which no points are removed are included in the supplemental material Fig. 2. 
\section{Discussion}\label{sec:discussion}
\par
Standard correlation has been widely used to analyze functional connectivity from rs-fMRI timeseries between prespecified ROIs. Prior work has shown its high sensitivity for detecting the existence of network architectures under both simulated and experimental scenarios~\citep{smith2011network,dawson2013evaluation}. This paper describes methodology for analyzing rs-fMRI data using a generalization of well-established correlation ideas. The generalization, denoted by ``p-correlation'' (``p'' for ``prediction''), is to compute the correlation between the $j^{th}$ signal and an optimal linear time-invariant causal estimate of the $j^{th}$ signal based on the $i^{th}$ signal. In this way, it captures additional features concerning the interaction between two ROIs, specifically, the causality and directionality of the information flow on which the interaction depends. Based on the finite-memory linear time-invariant causal model, p-correlation allows the memory duration to be different in the two directions for one pair of ROIs and also to be different for different pairs of ROIs. In contrast, structural vector autoregressive models \citep{kim2007unified,chen2011vector} are assumed to have the same memory duration across all ROIs. P-correlation is a generalization of standard correlation ideas because, if the estimate of the $j^{th}$ signal based on the $i^{th}$ signal is restricted to use only the current value of the $i^{th}$ signal, then p-correlation and standard correlation have the same magnitude.
\par 
Testing p-correlation on simulated fMRI data provided in~\cite{smith2011network}, the greater performance accuracy of p-correlation, which uses lagged information from the BOLD timeseries, demonstrates the importance of causal information which is missing in standard correlation. In our results, the mean duration of the impulse response estimated by $\AIC$ using a search limited to a maximum duration of 15s was roughly 4s. In these data, a search extending to 15s is not a restriction on the maximum duration. As is described in Table 1, the accuracy of p-correlation on the simulated data of Smith is about 0.5 (0.405 to 0.532). While higher levels are desirable, this performance exceeds the performance of many alternative algorithms on all four sets of simulations.
\par
Many approaches have been introduced to assess functional or effective connectivity of rs-fMRI data. \cite{smith2011network} evaluated the validity of 38 approaches~\citep[Fig.~4]{smith2011network} using simulated BOLD signals and a variety of performance measures. The methods tend to have different levels of performance for different measures, e.g., detection of a connection versus determination of the direction of a connection. The p-correlation approach introduced in this paper depends on causal dynamical models and so we focus on this particular aspect of previous work. Dynamic Causal Modeling (DCM) has been used with some success to assess causal dynamics in fMRI data by relying on sophisticated models of neural dynamics. As discussed in \citet[p.~878]{smith2011network}, most existing DCM algorithms require knowledge of external inputs (which are not known for rs-fMRI) although some variations may not~\citep{DaunizeauFristonKiebelPhysicaD2009}; all versions tend to be mathematically poorly conditioned; and all versions fail to scale to networks with large numbers of ROIs which are necessary for experimental studies. In contrast, the p-correlation approach described in this paper scales similarly to a correlation approach for which hundreds of ROIs are not a challenge~\citep{nan2014pcorrelation}.
\par
Several versions of Granger causality analysis, based on multivariate vector autoregressive modeling, have been tested and performed poorly~\citep{smith2011network}. Granger causality relies on regression and comparison of two predictions. The first prediction is based purely on an autoregressive model of the signal at the $i$th ROI based on the past of the same signal. The second prediction is based on regression of the signal at the $i$th ROI based on the past of the signal at the $j$th ROI and, possibly, an autoregression as in the first case. The sample covariances of the prediction errors are then combined, essentially by taking the ratio of the sample covariances scaled by integers describing the amounts of data, to yield a statistic that is distributed according to the Fisher-Snedecor $F$ distribution.
This statistic, indexed by $i$ and $j$, is used to fill an asymmetric matrix. Although both are based upon lagged information there are important differences between p-correlation and Granger causality. P-correlation is not a statistic comparing two possible dependencies but rather is a statistic measuring the accuracy of prediction using a particular dependency. The motivation for the Granger causality statistic is dependent on the original Gaussian assumptions on the errors when linear regression is used to describe the ROI time series. P-correlation is based on just the sample variance of the prediction error and does not have a Gaussian motivation which is advantageous if the BOLD signals lack Gaussian structure. Multivariate autoregressive processes have been used as the basis for generative models for complete sets of ROIs. Such models, which focus on the effect of the past on the present, can be combined with structural equation modeling (SEM) models, which focus on contemporaneous effects~\citep{chen2011vector}. 
\par 
Multivariate autoregressive processes (MVAR) have been successfully used in neuroscience outside of fMRI, e.g., in order to describe signals from EEG experiments \citep{kus2004determination,blinowska2009multivariate,ding2000short,babiloni2005estimation,ligeza2016interactions,korzeniewska2011dynamics,wilke2008estimation}. Both MVAR, e.g., Eq. 1 in \cite{kus2004determination}, and the linear regression model used in this paper (Eq. \ref{eq:LSconst2}) are regression models which predict one timeseries from either all timeseries which include oneself (MVAR) or from the past of another timeseries (Eq. \ref{eq:LSconst2}). Both predictions are characterized by impulse responses. The method introduced in \cite{kus2004determination} determines the connection strength based on the impulse response, whereas p-correlation determines the functional connectivity based on both the impulse response and the original timeseries. Existing literature, e.g., \cite{valdes2005estimating} and \cite{davis2015sparse}, has shown the robust estimation of the MVAR model by introducing sparse regression techniques, and the success of estimating functional connectivity through the sparse MVAR models. In addition, a conditional MVAR model, e.g., Ch 17.3 in \cite{schelter2006handbook}, may also be used to address the common driver problem. Other approaches to examining BOLD signal propagation using lags, as is done in p-correlation, have been highly reproducible~\citep{mitra2015lags}. In this paper, a linear regression model (Eq. \ref{eq:LSconst2}) is used as the predictor in p-correlation to estimate the causal relation between a pair of BOLD signals. Other lag-based predictors, e.g., MVAR based models, can also be adapted into the p-correlation concept, however, they would not have the result that duration of 1 sample (e.g., no lags) gives standard correlation.
\par 
In addition to the algorithms used in~\cite{smith2011network}, which estimate the directional connectivity for single subject data sets, the IMaGES \citep{ramsey2010six,ramsey2011multi} and GIMME~\citep{gates2012group} algorithms use a group of subjects. While these algorithms provide better performance in situations where groups of subjects can be analyzed collectively, both algorithms have challenges. The sparse graph estimated by IMaGES for a group of subjects does not tell the strengths of the connectivity and ``will not reflect the variation of a group''~\citep[p.571]{mumford2014bayesian}. Similar to DCM's limitation on scalability, small networks with less than 25 ROIs are well analyzed by the GIMME algorithm. However, its performance on large-scale functional networks is not known. As p-correlation can work with hundreds of ROIs, it can be used in evaluating large-scale brain networks. Furthermore, p-correlation can work on individual subjects so it potentially could be applied to patient clinical data. Other algorithms that estimate direction after a connection is already detected also exist (Section \ref{sec:simulated:competitors}). While such algorithms may be useful in some circumstances, they do not allow for situations where both directions are present but of different strengths.
\par
The \cite{smith2011network} simulated data has lower dimensionality than experimental brain data. For instance, in the simulation, connections are all unidirectional while most neural connections are bidirectional. Additionally, in the simulations, most connections had a value of exactly zero. Furthermore, it introduces unrealistic noise and it has a large number of parameters that must be set and which influence the resulting simulation \citep{wang2014systematic}. 
While the \cite{smith2011network} simulated data is not completely realistic and it is not a perfect test of p-correlation, this data continues to be used \citep{smith2011network,gates2012group,ramsey2011multi,hyvarinen2013pairwise,ryali2016multivariate}, and the results continue to be discussed \citep{geerligs2016functional}. In this paper, we leveraged the same data used in~\cite{smith2011network} for comparison with other published metrics, providing a broader context for these findings. We hope to use a broader range of simulated data to further validate p-correlation in our future work.
\par
In order to focus on the challenges of a ``common driver'', we have produced additional synthetic data for the three ROI network of Fig. \ref{fig:commondriver} in which one ROI drives two other ROIs but the two other ROIs do not directly interact. Using p-correlation in this network we found that p-correlation can identify the existence and direction of the interactions between the driving ROI and the other two ROIs (even when the two interactions are of different strengths). Furthermore, p-correlation did not introduce false interactions between the two driven ROIs.
\par 
We have applied p-correlation to experimental data from the 1000 Functional Connectome Project \citep{biswal2010toward}. The p-correlation approach successfully replicated the modular architecture of the local and distributed networks previously reported using standard correlation~\citep{nan2014pcorrelation}
(see Section \ref{sec:ExperimentalPerformance} Fig. \ref{fig:Stability_BIC}). Highly correlated p-correlation values on the two different cohorts also demonstrated that the p-correlation is highly reproducible and thus robust on experimental data. A current limitation of the p-correlation approach is that missing nodes cannot be accommodated, thereby limiting an extension of this approach to lesioned populations.
\par
Here we introduce a novel concept, the p-correlation, to estimate brain connectivity within well-characterized large-scale functional networks. The replication of previously observed network architectures in experimental data and the performance against the ground truth in simulated data, both suggest that the p-correlation approach may hold promise for future investigations of the brain's dynamic functional architecture.

\section{Acknowledgements}\label{sec:ackowledgements}
We are very grateful to Prof. S.M. Smith (University of Oxford) for providing simulation data and his software for applying “Patel's conditional dependence measures” and network measurements as described in his paper~\citep{smith2011network} as well as for helpful discussion. We are also very grateful to Dr. Chandler Lutz for providing the pairwise Granger causality code, Dr. Rodrigo Quian Quiroga for providing the generalized synchronization code, and Drs. Shohei Shimizu, Patrik Hoyer and Aapo Hyvärinen for providing LiNGAM/FastICA. Nan Xu and Peter C. Doerschuk are grateful for support by NSF Grant 1217867.
\clearpage
\pagenumbering{gobble}
\section*{Supplemental}
\begin{figure}[h!]
	\hspace*{.05in}
	\centerline{\psfig{figure=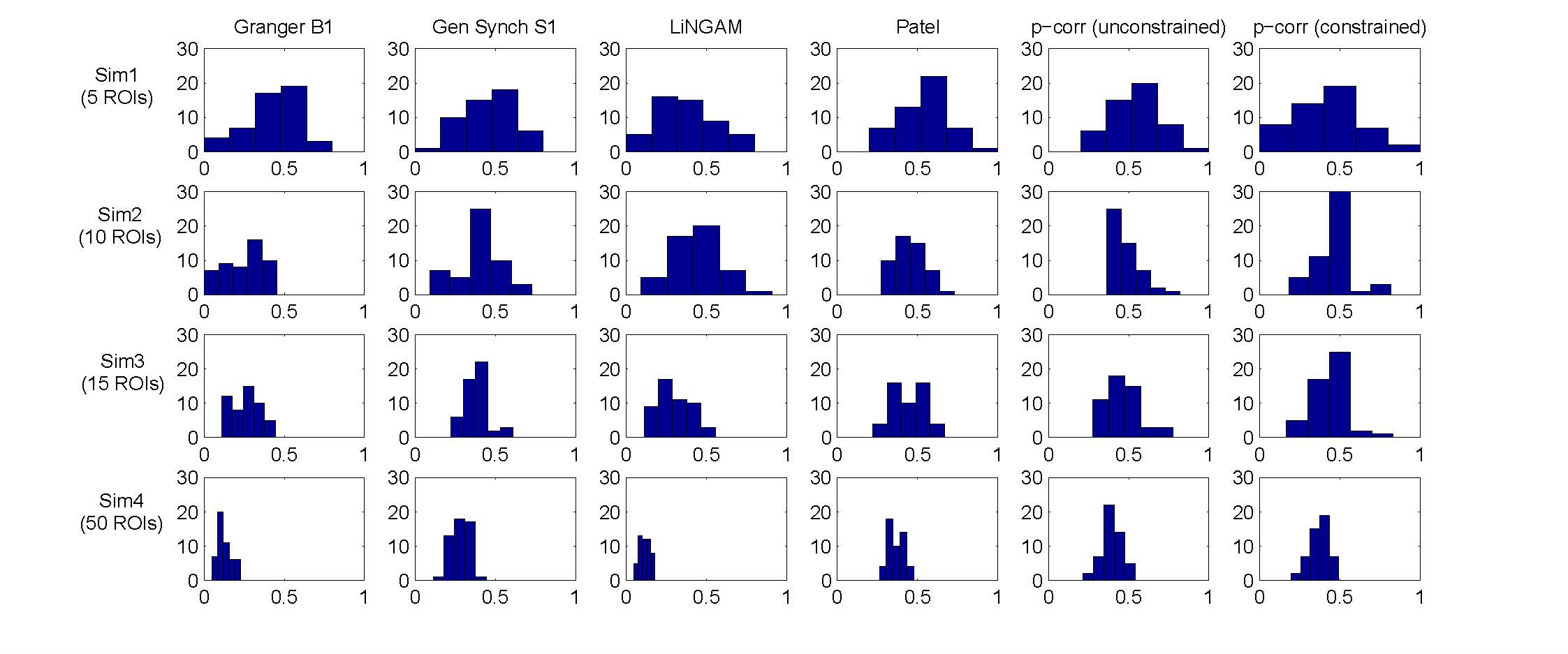, height=80mm} }
	\vspace*{-.3in}
	\caption*{
		\label{fig:Histogram2}
		Figure S1: Accuracy histogram for Granger B1, Gen Synch S1, LiNGAM, Patel and  p-correlation  with unconstrained and constrained Least Squares.
		}
\end{figure}
\begin{figure}[h!]
	\begin{center}
		\begin{tabular}{cc}
			\includegraphics[width=5cm]{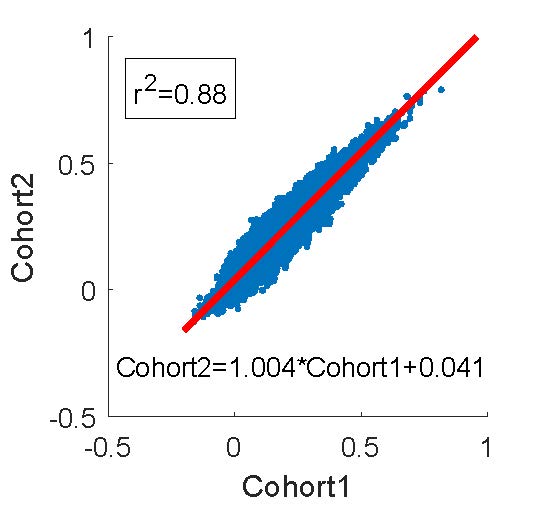}
			&
			\includegraphics[width=5cm]{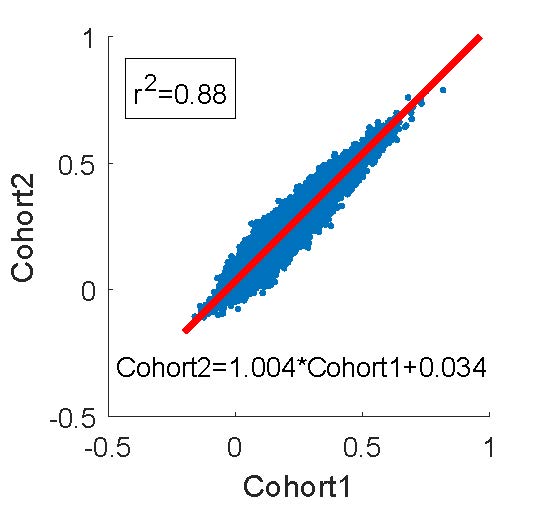}
			\\
			(a) p-correlation
			&
			(b) correlation 
		\end{tabular}	
	\end{center}
	\caption*{\label{fig:Scatter_corr_pcorrAll}
		Figure S2: Scatter plot of p-correlation and correlation for the two cohorts. The red line is the Least Squares fit for predicting Cohort 2 from Cohort 1. 
		}
\end{figure}
\clearpage		

\bibliography{mybibfileFINAL}

\end{document}